# AIR: fused Analytical and Iterative Reconstruction method for computed tomography

Liu Yang
*Department of Radiology and Imaging Sciences, Emory University, Atlanta, Georgia 30322;*
*School of Biomedical Engineering, Southern Medical University, Guangzhou, China 510515*
Yu Gao
*Department of Mathematics and Computer Science, Emory University, Atlanta, Georgia 30322*
Sharon X. Qi
*Department of Radiation Oncology, University of California, Los Angeles, California 90095*
Hao Gao[a]
*School of Biomedical Engineering, Shanghai Jiao Tong University, Shanghai, China 200240;*
*Departments of Mathematics and Computer Science, and Radiology and Imaging Sciences, Emory University, Atlanta, Georgia 30322*

**Purpose:** CT image reconstruction techniques have two major categories: analytical reconstruction (AR) method and iterative reconstruction (IR) method. AR reconstructs images through analytical formulas, such as filtered backprojection (FBP) in 2D and Feldkamp-Davis-Kress (FDK) method in 3D, which can be either mathematically exact or approximate. On the other hand, IR is often based on the discrete forward model of X-ray transform and formulated as a minimization problem with some appropriate image regularization method, so that the reconstructed image corresponds to the minimizer of the optimization problem. This work is to investigate the fused analytical and iterative reconstruction (AIR) method.
**Methods:** Based on IR with L1-type image regularization, AIR is formulated with a AR-specific preconditioner in the data fidelity term, which results in the minimal change of the solution algorithm that replaces the adjoint X-ray transform by the filtered X-ray transform. As a proof-of-concept 2D example of AIR, FBP is incorporated into tensor framelet (TF) regularization based IR, and the formulated AIR minimization problem is then solved through split Bregman method with GPU-accelerated X-ray transform and filtered adjoint X-ray transform.

**Results:** FBP, IR, and AIR were compared using Siemens imaging quality phantom scanned with TomoTherapy megavoltage CT. The reconstruction was performed with the regular fully sampled case (100% data) and the low-dose undersampled case (25% data). Quantitative contrast and resolution evaluations were computed, including the full width at half maximum (FWHM) and the contrast-to-noise ratio (CNR). The experimental results suggest that AIR provides better image resolution and contrast than FBP or IR.

**Conclusion:** AIR, the fused Analytical and Iterative Reconstruction method, is proposed with a proof-of-concept 2D example to synergize FBP and TF-regularized IR, with improved image resolution and contrast for experimental data. The potential impact of AIR is that it offers a general framework to develop various AR enhanced IR methods, when neither AR nor IR alone is sufficient.

Key words: CT image reconstruction, iterative reconstruction, analytical reconstruction, low-dose, tensor framelet, GPU

**Introduction**

CT image reconstruction techniques[1] can be classified roughly into two categories: analytical reconstruction (AR) method and iterative reconstruction (IR) method. Inspired by compressed sensing[2-3], L1-type IR techniques[4-14] for computed tomography (CT) have been well studied in the last decade, which provide the improved imaging quality over AR in terms of noise and artifact reduction and contrast enhancement, particularly for low-dose CT. However, the image resolution from IR may sometimes be compromised[15], especially with the noisy data from low-dose or partial-view scans, while AR maintains the resolution despite of noise and streaking artifacts. On the other hand, the conventional wisdoms in AR[16-22] seem to be irrelevant for IR.

This work is to investigate the fused analytical and iterative reconstruction (AIR) method. We will present a novel mathematical framework to fuse AR and IR, study a 2D example of AIR by combining FBP and tensor framelet (TF) regularized IR[13,15], and perform the proof-of-concept experimental studies using the TomoTherapy (Accuray Inc., Sunnyvale, CA) megavoltage CT scans of a Siemens imaging quality phantom (Siemens Medical Solution, Concord, CA)[23].

**Methods**

The key of AIR for integrating AR and IR is to consider the preconditioned version of the IR formulation, i.e.,

$$X = \arg\min_X \| P(AX - Y) \|_2^2 + \lambda R(X), \qquad (1)$$

where $P$ comes from a AR-specific filtration step that will be explained next. In Eq. (1), $A$, $X$ and $Y$ stand for the system matrix, the unknown image, and the projection data, respectively. $R(X)$ is the image regularization term with the balancing parameter $\lambda$.

Most AR methods contain or approximately have two steps: first filter the data and then backproject the filtered data. Let $A^{-1}$ denote the AR method, then $A^{-1}=BC$ with the backprojection operator $B$ and filtration operator $C$. Then let us define $P=C^{1/2}$, in the sense that

$$C = (F^T \hat{C}^{1/2} F)^T \cdot (F^T \hat{C}^{1/2} F) = (C^{1/2})^T \cdot C^{1/2}, \qquad (2)$$

where $F$ is the Fourier transform. Eq. (2) utilizes the fact that the filtration operator $C$ is diagonalizable in the Fourier domain. Then we have

$$P = P^T = C^{1/2}. \qquad (3)$$

As an illustration, let us consider AIR with a simple L2 regularization $R(X)=\|X\|^2$. From the optimal condition of Eq. (1), the solution can be obtained by solving

$$(A^T P^T PA + \lambda I)X = A^T P^T PY. \qquad (4)$$

Next, by introducing the filtered adjoint X-ray transform operator $A^T = A^T P^T P$ and using Eq. (2) and (3), we have

$$(A'^T A + \lambda I)X = A'^T Y. \quad (5)$$

That is, due to the preconditioner $P$, the adjoint X-ray transform $A^T$ is replaced by the filtered adjoint X-ray transform $A'^T$, which is slightly different from the AR operator $A^{-1}$.

This feature with the minimal change in solution algorithm also applies to L1-type regularization. For example, with the Split Bregman method[24] as the solution algorithm[10-11,15], the major change from IR to AIR is the L2 step in the Bregman loop, which can be solved in the similar fashion as Eq. (5). As an another example, when using the forward-backward operator splitting method[25], the proximal step is the same for IR and AIR, and the forward step becomes

$$X^{k+1/2} = X^k - \mu A'^T(AX^k - Y). \quad (6)$$

Again, the only difference in solution algorithm for IR and AIR is that $A^T$ is replaced by $A'^T$, with a filtration step followed by the adjoint X-ray transform.

**Materials**

In this proof-of-concept study, we implemented a 2D version of AIR using tensor framlet regularization[13,15]. FBP was used as the AR method, and the Split Bregman method[10-11,15,24] was used for solving IR and AIR. The reconstruction codes were parallelized on GPU with the fast parallel algorithms[26] for computing $A$, $A^T$ and $A'^T$.

The experimental data was acquired using a Siemens image quality phantom from a TomoTherapy HD unit with on-board megavoltage CT imaging system. There were 528 effective detection pixels per view, and 800 or 200 projection views per rotation for the fully-sampled case (100% data) or the equally-undersampled case (25% data). The images were reconstructed to a 350×350 square pixel array with 1×1 mm² resolution. The same imaging geometric parameters[15] were used for FBP, IR, and AIR.

**Results**

The resolution slice is displayed in Fig. 1 (100% data) and Fig. 2 (25% data) with the quantitative FWHM values in Table 1 and the Cross-line plots in Fig. 5. With 100% data, suggested by Figs. 1(a) and 1(c), AIR provided better image resolution than FBP, which was due to the use of TF image regularization for noise suppression. From Figs. 1(b) and 1(c), AIR also had better image resolution than IR, which was due to the fused FBP step in IR for preventing over-smoothing. For example, the bar group with 4 line pairs per cm (lp/cm) was clearly better distinguishable for AIR (Fig. 1(f)) than for either FBP (Fig. 1(d)) or IR (Fig. 1(e)). The same observation carried over to 25% data, suggested by Fig. 2. For example, the bar group with 3lp/cm was clearly better distinguishable for AIR (Fig. 2(f)) than for either FBP (Fig. 2(d)) or IR (Fig. 2(e)). Moreover, the superiority of AIR in imaging resolution was also confirmed from the profile plots of the 3lp/cm bars in Fig. 5 and their corresponding full-width-at-half-maximum (FWHM) values.

The contrast slice is displayed in Fig. 3 (100% data) and Fig. 4 (25% data) with the quantitative contrast-to-noise (CNR) values in Table 2 (i.e., $|\mu_t-\mu_b|/(\sigma_t^2+\sigma_b^2)^{0.5}$ with the averaged value $\mu_t/\mu_b$ of the target/background, and the standard deviation $\sigma_t/\sigma_b$ of the target/background). With 100% data, suggested by Figs. 3(a) and 3(c) and Table 2, AIR had better image contrast than FBP, which was again due to the use of TF image regularization for noise suppression. Furthermore, suggested by Figs. 3(b) and 3(c) and Table 2, AIR also had slightly better image contrast than IR. With 25% data, suggested by Fig. 4 and Table 2, the same situation was observed.

**Conclusion**

We have proposed AIR, a novel mathematical image reconstruction framework, to integrate AR and IR, and demonstrated the improved imaging quality of a proof-of-concept 2D version of AIR from AR or IR using experimental data scanned from TomoTherapy megavoltage CT with either fully-sampled data or equally-undersampled data.

The potential impact of AIR is that it offers a general framework to utilize the conventional wisdoms from both AR and IR, and develop AR enhanced IR under various settings with different utilities, when either AR or IR alone is insufficient.

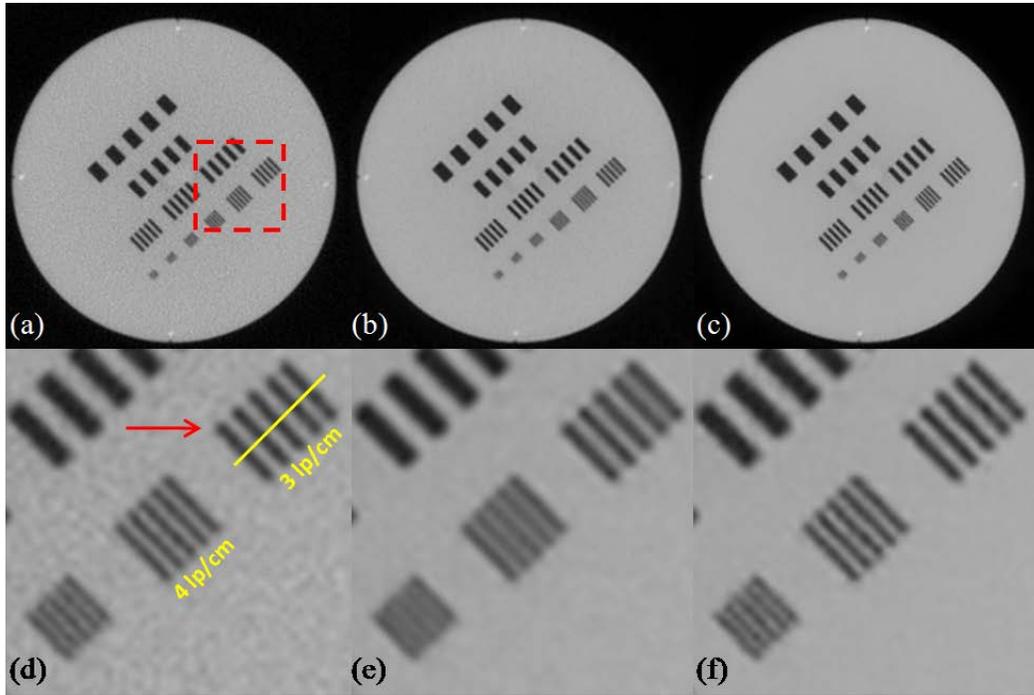

FIG. 1. The evaluation of image resolution with 100% data. (a)-(c) are from FBP, IR, and AIR respectively, and their corresponding zoom-in details are in (d)-(f).

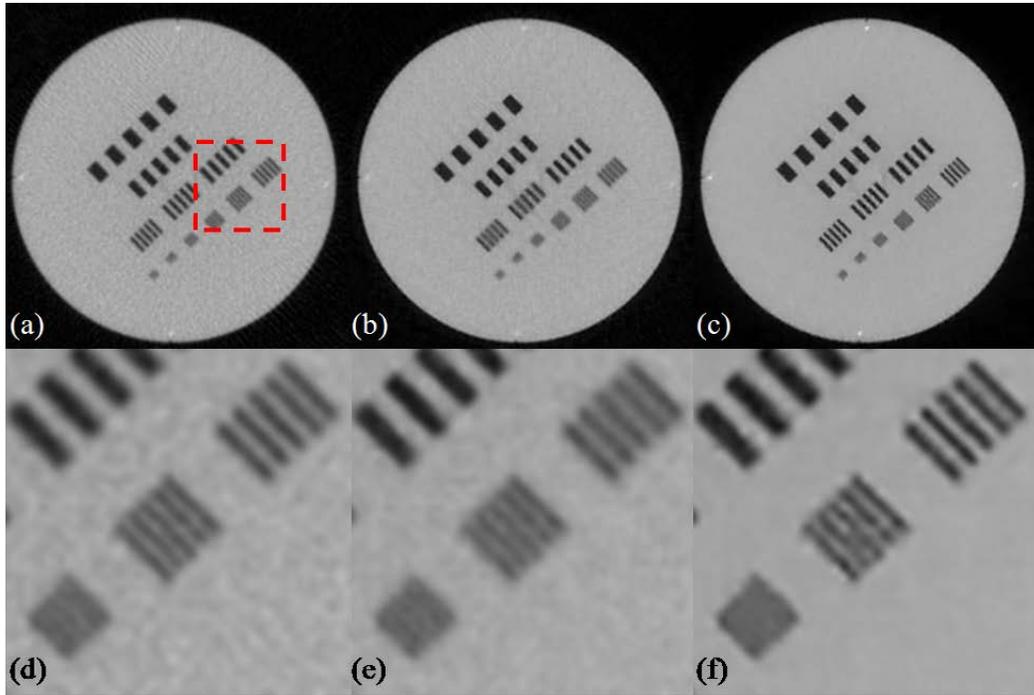

155

FIG. 2. The evaluation of image resolution with 25% data. (a)-(c) are from FBP, IR, and AIR respectively, and their corresponding zoom-in details are in (d)-(f).

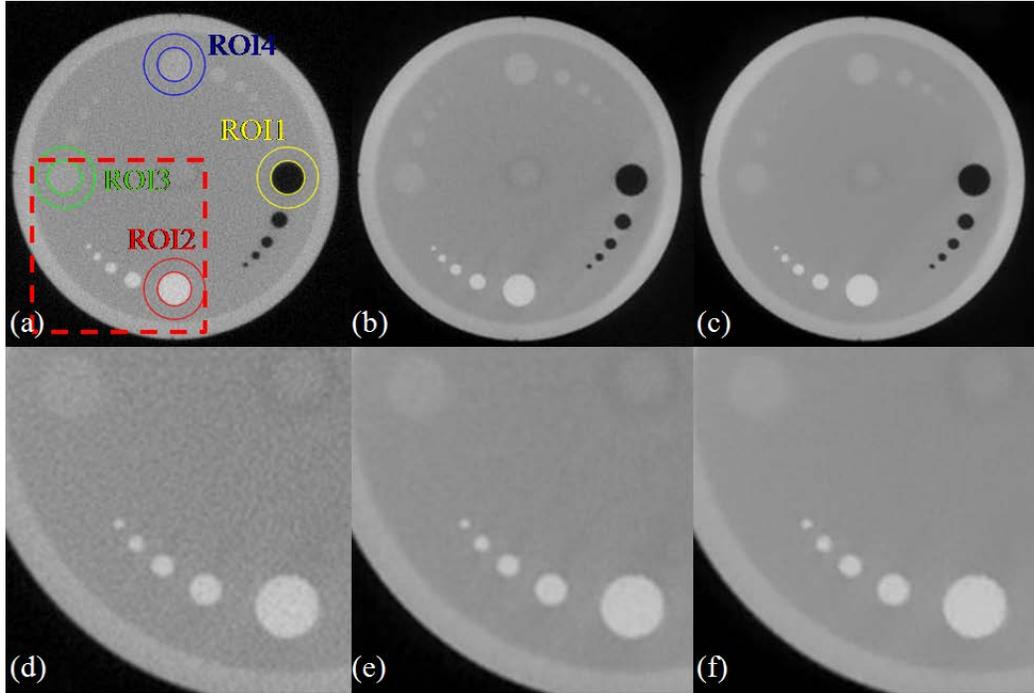

FIG. 3. The evaluation of image contrast with 100% data. (a)-(c) are from FBP, IR, and AIR respectively, and their corresponding zoom-in details are in (d)-(f). Here the ROI's for quantitative CNR evaluations are marked in (a) with each smaller circle for the target and annulus as the background for the CNR calculation.

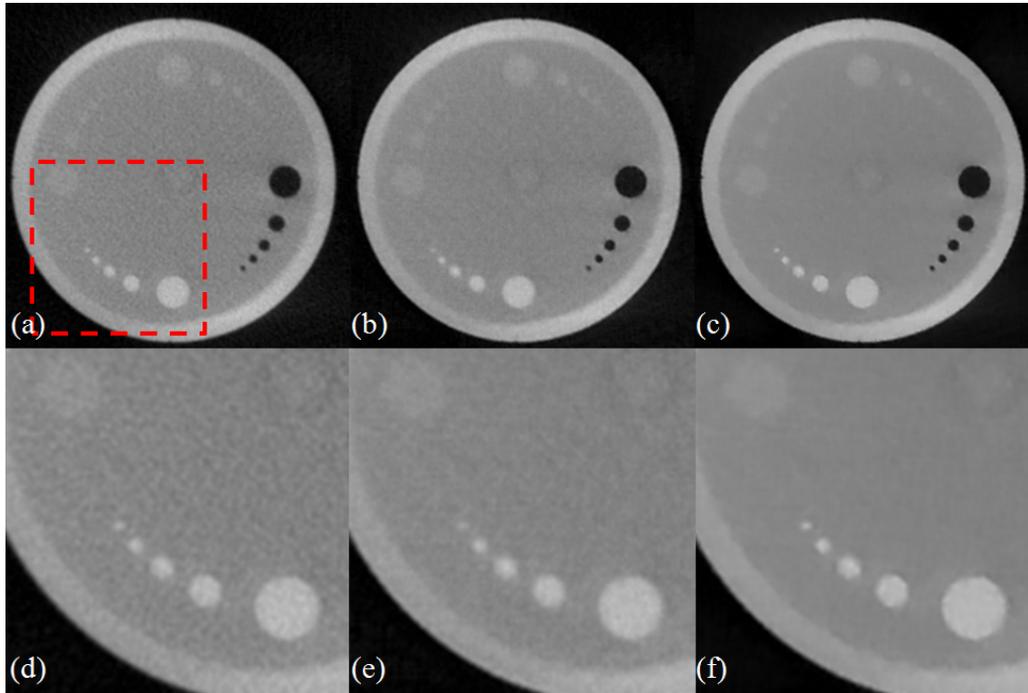

FIG. 4. The evaluation of image contrast with 25% data. (a)-(c) are from FBP, IR, and AIR respectively, and their corresponding zoom-in details are in (d)-(f).

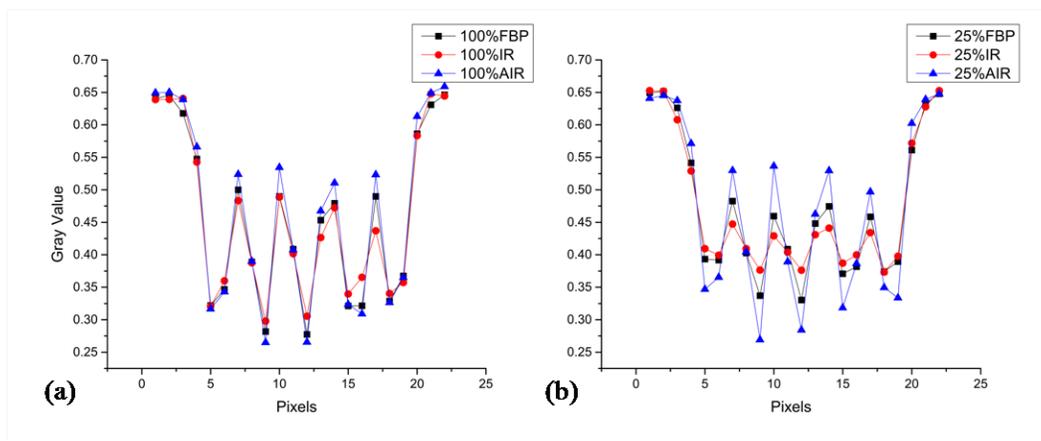

FIG. 5. The profiles of 3lp/cm bars(Fig. 1(d)) of the resolution slice. (a): 100% data; (b): 25% data.

TABLE. 1. Quantitative resolution evaluation via the FWHM values of the 3lp/cm bars (Fig. 1(d)).

|  | FBP | IR | AIR |
|---|---|---|---|
| 100% data | 1.76 | 1.92 | 1.66 |
| 25% data | 1.92 | 1.99 | 1.75 |

TABLE. 2. Quantitative contrast evaluation via the CNR values of ROI's (Fig 3(a))

|  |  | ROI1 | ROI2 | ROI3 | ROI4 |
|---|---|---|---|---|---|
| 100% data | FBP | 4.32 | 3.40 | 1.43 | 1.03 |
|  | IR | 4.53 | 4.00 | 2.38 | 2.05 |
|  | AIR | 4.59 | 4.03 | 2.54 | 2.26 |
| 25% data | FBP | 4.04 | 3.42 | 1.41 | 1.18 |
|  | IR | 4.17 | 3.46 | 1.58 | 1.43 |
|  | AIR | 4.32 | 3.85 | 1.99 | 2.07 |

[a] Electronic mail: hao.gao.2012@gmail.com
[1] T. M. Buzug, Computed Tomography: From Photon Statistics to Modern Cone-Beam CT. Springer, (2008).
[2] E. J. Candès, J. Romberg, and T. Tao, "Robust uncertainty principles: exact signal reconstruction from highly incomplete frequency information", IEEE Trans. Inf. Theory **52**, 489-509 (2006).
[3] D. L. Donoho, "Compressed sensing", IEEE Trans. Inf. Theory, **52**, 1289-306 (2006).
[4] E. Y. Sidky, C.-M. Kao, and X. Pan, "Accurate image reconstruction from few-views and limited-angle data in divergent-beam CT," J. X-Ray Sci. Technol. **14**, 119-139 (2006).
[5] J. Wang, T. Li, H. Lu, and Z. Liang, "Penalized weighted least-squares approach to sinogram noise reduction and image reconstruction for low-dose X-ray computed tomography," IEEE Trans. Med. Imaging **25**, 1272-1283 (2006).
[6] G. H. Chen, J. Tang, and S. Leng, "Prior image constrained compressed sensing (PICCS): a method to accurately reconstruct dynamic CT images from highly undersampled projection data sets," Med. Phys. **35**, 660-663 (2008).
[7] H. Y. Yu and G. Wang, "Compressed sensing based interior tomography, " Phys. Med. Biol. **54**, 2791-2805 (2009).
[8] K. Choi, J. Wang, L. Zhu, T.-S. Suh, S. Boyd, and L. Xing, "Compressed sensing based cone-beam computed tomography reconstruction with a first-order method," Med. Phys. **37**, 5113-5125, (2010).

[9] X. Jia, B. Dong, Y. Lou, and S. B. Jiang, "GPU-based iterative cone-beam CT reconstruction using tight frame regularization," Phys. Med. Biol. **56**, 3787-3806, (2010).

[10] H. Gao, J. F. Cai, Z. Shen, and H. Zhao, "Robust principal component analysis-based four-dimensional computed tomography," Phys. Med. Biol. **56**, 3181-3198 (2011).

[11] H. Gao, H. Yu, S. Osher, and G. Wang, "Multi-energy CT based on a prior rank, intensity and sparsity model (PRISM)," Inverse Problems **27**, 115012 (2011).

[12] T. Niu and L. Zhu, "Accelerated barrier optimization compressed sensing (ABOCS) reconstruction for cone-beam CT: Phantom studies," Med. Phys. **39**, 4588-4598 (2012).

[13] H. Gao, R. Li, Y. Lin, and L. Xing, "4D cone beam CT via spatiotemporal tensor framelet," Med. Phys. **39**, 6943-6946 (2012).

[14] B. Zhao, H. Gao, H. Ding, and S. Molloi, "Tight-frame based iterative image reconstruction for spectral breast CT, " Med. Phys. **40**, 031905 (2013).

[15] H. Gao, X. S. Qi, Y. Gao, and D. A. Low, "Megavoltage CT imaging quality improvement on TomoTherapy via tensor framelet," Med. Phys. **40**, 081919 (2013).

[16] L. A. Feldkamp, L. C. Davis, and J. W. Kress, "Practical cone-beam algorithm," J. Opt. Soc. Am. A. **1**, 612-619 (1984).

[17] G. Wang, T. H. Lin, P. C. Cheng, and D. M. Shinozaki, "A general cone-beam reconstruction algorithm, " IEEE Transaction on Medical Imaging **12**, 486-496 (1993).

[18] K. C. Tam, S. Samarasekera, and F. Sauer, "Exact cone beam CT with a spiral scan," Phys. Med. Biol. **43**, 1015-1024 (1998).

[19] H. Kudo, F. Noo and M. Defrise, "Cone-beam filtered-backprojection algorithm for truncated data," Phys. Med. Biol. **43**, 2885-2909 (1998).

[20] A. Katsevich, "Theoretically exact FBP-type inversion algorithm for spiral CT," SIAM J. Appl. Math. **62**, 2012-2026 (2002).

[21] G. H. Chen, "An alternative derivation of Katsevich's cone-beam reconstruction formula," Med. Phys. **30**, 3217-3226 (2003).

[22] Y. Zou and X. Pan, "Exact image reconstruction on PI-lines from minimum data in helical cone-beam CT," Phys. Med. Biol. **49**, 941-959 (2004).

[23] O. Gayou and M. Miften, "Commissioning and clinical implementation of a mega-voltage cone beam CT system for treatment localization," Med. Phys. **34**, 3183-3192 (2007).

[24] T. Goldstein and S. Osher, "The split Bregman algorithm for $l_1$ regularized problems," SIAM J. Imaging Sci. **2**, 323-343 (2009).

[25] P. L. Combettes and V. R. Wajs, "Signal recovery by proximal forward-backward splitting," Multiscale Model. Simul. **4**, 1168-1200 (2005).

[26] H. Gao, "Fast parallel algorithms for the x-ray transform and its adjoint," Med. Phys. **39**, 7110-7120 (2012).